\documentclass[twocolumn,preprintnumbers,amsmath,amssymb,superscriptaddress,10pt]{revtex4}
\usepackage{amssymb}
\usepackage{graphicx,color}
\usepackage{bm}
\usepackage{ulem}
\usepackage{color}

\begin{document}

\title{Viscous effects on the dynamical evolution of QCD matter during the first-order
confinement phase transition in heavy-ion collisions}

\author{Bohao Feng}
\affiliation{Department of Physics, Tsinghua University and Collaborative Innovation Center of Quantum Matter, Beijing 100084, China}

\author{Carsten Greiner}
\affiliation{Institut f$\ddot{u}$r Theoretische Physik, Johann Wolfgang Goethe-Universit$\ddot{a}$t Frankfurt, Max-von-Laue-Strasse 1, 60438 Frankfurt am Main, Germany}

\author{Shuzhe Shi}
\affiliation{ Physics Department and Center for Exploration of Energy and Matter, Indiana University, 2401 N Milo B. Sampson Lane, Bloomington, IN 47408, USA}

\author{Zhe Xu \footnote{xuzhe@mail.tsinghua.edu.cn}}
\affiliation{Department of Physics, Tsinghua University and Collaborative Innovation Center of Quantum Matter, Beijing 100084, China}

\begin{abstract}
We investigate viscous effects on the dynamical evolution of QCD matter during the
first-order phase transition, which may
happen in heavy-ion collisions. We first obtain the first-order phase transition line
in the QCD phase diagram under the Gibbs condition by using the MIT bag model
and the hadron resonance gas model for the equation of state of partons and hadrons. 
The viscous pressure, which corresponds to the friction in the energy balance,
is then derived from the energy and net baryon number conservation
during the phase transition. We find that the viscous pressure relates to the
thermodynamic change of the two-phase state and thus affects the timescale of the
phase transition. Numerical results are presented for demonstrations.
\end{abstract}

\maketitle

\section{Introduction}
\label{sec1:intro}
A phase diagram separates phases and determines conditions, at which different phases
coexist at thermal equilibrium. The completion of the QCD phase diagram 
\cite{Mohanty:2009vb} is an ongoing task and essential for understanding the matter under
strong interaction. Lattice QCD calculations \cite{Brown:1990ev} showed that the QCD
phase transition at small baryon chemical potential is a crossover rather than a real phase
transition. At high baryon chemical potential, theory predicts a first-order phase transition
line \cite{Asakawa:1989bq} ending at a QCD critical point  \cite{Stephanov:2004wx}.
The phase transition of QCD matter can be investigated in experiments of heavy-ion
collisions, where quark-gluon plasma (QGP) cools down due to expansion and hadronizes
at certain temperature and baryon chemical potential. One major goal of the beam energy
scan program at Relativistic Heavy Ion Collider (RHIC) 
\cite{Aggarwal:2010wy,Aggarwal:2010cw,Kumar:2012fb,Luo:2015doi} is to locate the
critical point in the QCD phase diagram.

Usually, a phase transition is defined when two coexisting phases are in thermal equilibrium.
The QCD matter produced in heavy-ion collisions possesses, however,  a nonzero viscosity
\cite{Romatschke:2007mq,Xu:2007jv,Song:2008hj} and deviates from thermal equilibrium.
If the system is not far away from  thermal equilibrium, one can still define thermodynamic 
quantities such as temperature, pressure, and chemical potential, as done in viscous
hydrodynamic calculations. With these thermodynamic quantities one can also identify
the first-order phase transition for the expanding QGP, if the Gibbs condition holds.
In this paper we consider nonzero viscosities of QCD matter and investigate viscous
effects on the dynamical evolution of QCD matter during the first-order phase transition.

In Sec. \ref{sec2:phase} we first calculate the first-order phase transition line under
the Gibbs condition for phase equilibrium by using MIT bag model and hadron resonance
gas model for the equation of state (EoS) of the parton and hadron phase.
We then show in Sec. \ref{sec3:viscous} how the shear and bulk viscosity affect the 
phase transition of the QCD matter produced in heavy-ion collisions.
Two different expansion geometries are applied to the evolution of the QCD matter.
In Sec. \ref{sec4:lim} further discussions are given.

\section{Phase Diagram}
\label{sec2:phase}
Since lattice QCD results of EoS at finite baryon chemical potential are not yet
available, we use the EoS from model calculations. Based on these, we present
in this section the first-order phase transition line in the temperature-baryon chemical
potential diagram.

The parton phase is considered as a system of massless quarks and gluons,
which interactions are described by perturbative QCD (pQCD) up to $g^2_s$
terms \cite{Waldhauser:1989ec,Shuryak:1980tp,Singh:2009jd}.
The pressure and energy density are
\begin{eqnarray}
\label{pressureP}
P_p&=&a(T,\mu_q,g_s)T^4-B \,, \\
\label{edensityP}
e_p&=&3a(T,\mu_q,g_s)T^4+B\,,
\end{eqnarray}
where $B$ is the bag constant with $B^{1/4}=200 \mbox{ MeV}$ and
\begin{eqnarray}
&&a(T,\mu_q,g_s) \nonumber \\
&=&\frac{\pi^2}{45} \left [ 8+\frac{21}{4}n_f+\frac{45}{2\pi^2} \sum^{n_f}_{i=1}
\left ( \frac{\mu^2_i}{T^2}+\frac{\mu^4_i}{2\pi^2T^4} \right ) \right ] \nonumber \\
&& -\frac{8}{144}g^2_s \left [ 3+\frac{5}{4}n_f+\frac{9}{2\pi^2}\sum^{n_f}_{i=1}
\left ( \frac{\mu^2_i}{T^2}+\frac{\mu^4_i}{2\pi^2T^4} \right ) \right ] \,.
\end{eqnarray}
$\mu_i$ is the chemical potential of a quark flavor. $n_f$ is the number of
quark flavors. We consider $u,d,s$ quarks $(n_f=3)$ and assume 
$\mu_u=\mu_d \equiv \mu_q$, $\mu_{\bar u}=\mu_{\bar d}=- \mu_q$,
and $\mu_s=\mu_{\bar s}=0$.
The running coupling is given by \cite{Waldhauser:1989ec,Shuryak:1980tp,Singh:2009jd}.
\begin{equation}
\alpha_s=\frac{g^2_s}{4\pi}=\frac{12\pi}{33-2n_f} 
\left ( \ln \frac{0.8\mu_q^2+15.6T^2}{\Lambda_{QCD}^2} \right )^{-1}
\end{equation}
with $\Lambda_{QCD}=100 \mbox{ MeV}$. From the pressure we obtain the net
baryon number density, which is one third of the net quark number density,
\begin{equation}
n_{Bp}=\frac{1}{3} \left.\frac{\partial P_p}{\partial \mu_q}\right|_{T}
\approx \frac{1}{3} n_f \left( 1-\frac{2\alpha_s}{\pi} \right ) 
\left ( \mu_qT^2+\frac{1}{\pi^2}\mu_q^3 \right ) \,.
\end{equation}
Here we neglect the logarithmic dependence of $\alpha_s$ on $\mu_q$. 

The hadron phase is described by the hadron resonance gas model (HRG) 
\cite{Kouno:1988bi,Singh:1994cy,Tiwari:2013wga}. Baryons, mesons, and
their resonances having masses up to $2 \mbox{ GeV}$ are included.
The pressure and energy density of hadrons and the net baryon number
density are given by
\begin{eqnarray}
\label{pressureH}
P_{h}&=&\frac{\sum_i P^0_i}{1+\sum_i n^0_i v_i}+\frac{\sum_j \bar{P}^0_j}{1+\sum_j \bar{n}^0_j v_j}
+\sum_m P^0_m \,, \\
\label{edensityH}
e_{h}&=&\frac{\sum_j e^0_i}{1+\sum_i n^0_i v_i}+\frac{\sum_j \bar{e}^0_j}{1+\sum_j \bar{n}^0_j v_j}
+\sum_m e^0_m \,, \\
\label{ndensityH}
n_{Bh}&=&\frac{\sum_i n^0_i}{1+\sum_i n^0_i v_i}
-\frac{\sum_j \bar{n}^0_j}{1+\sum_j \bar{n}^0_j v_j} \,,
\end{eqnarray}
where $i$, $j$, and $m$ denote baryon, antibaryon, and meson, respectively.
$P^0_k$, $e^0_k$, and $n^0_k$ are the pressure, energy density, and number
density of a hadron species $k$, when assuming a non-interacting hadron gas,
\begin{eqnarray}
P_k^0(T,\mu^h_k)&=&\frac{d_k}{6\pi^2}\int dp \frac{p^4}{\sqrt{p^2+m_k^2}}f_k(p) \,,\\
e_k^0(T,\mu^h_k)&=&\frac{d_k}{2\pi^2}\int dp p^2 \sqrt{p^2+m_k^2} f_k(p) \,, \\
n_k^0(T,\mu^h_k)&=&\frac{d_k}{2\pi^2}\int dp p^2 f_k(p) \,,
\end{eqnarray}
where $d_k$ is the degeneracy factor and
\begin{equation}
f_k(p)=\left [ \exp \left ( \frac{\sqrt{p^2+m_k^2}-\mu^h_k}{T} \right ) \pm 1 \right ]^{-1} \,.
\end{equation}
$+$ sign is for baryons and $-$ sign is for mesons. $m_k$ denotes the hadron
mass and  $\mu^h_k$ denotes the hadron chemical potential, which relates to the
quark chemical potential as $\mu^h_k=\nu_k\mu_q$, where $\nu_k$ is the net quark
number in hadron $k$. Thus, $\mu^h_k=0$ for mesons and  
$\mu^h_{\bar {k}}=-\mu^h_k$ when $\bar {k}$ denotes the antiparticle of $k$.

We take short range repulsive interactions among (anti)baryons into account 
\cite{Kouno:1988bi,Cleymans:1986cq,Rischke:1991ke,BraunMunzinger:1999qy}.
This is indicated in the denominators of the terms in Eqs. (\ref{pressureH})-(\ref{ndensityH}),
where $v_{i(j)} =4\pi r_{i(j)}^3/3$ is the eigen volume of baryon $i$ or antibaryon $j$.
We assume a same hard sphere radius for all baryons.

\begin{figure}[ht]
 \centering
 \includegraphics[width=0.45\textwidth]{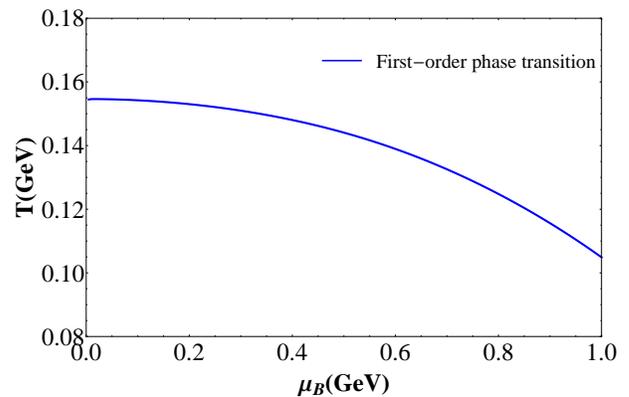}
 \caption{The QCD first-order phase transition line from model calculations.
}
 \label{fig1-phase}
\end{figure}

During the first-order phase transition the Gibbs condition holds. The pressure,
temperature and baryon chemical potential of both parton and hadron phase are
equal. Equating the pressures from Eqs. (\ref{pressureP}) and (\ref{pressureH})
with the same temperature $T$ and baryon chemical potential $\mu_B=3\mu_q$,
we obtain the phase boundary curve, which is shown in Fig. \ref{fig1-phase}.
The hard-core radius $r_{i(j)}$ of (anti)baryons is set to be 
$0.6 \mbox{ fm}$ \cite{Kouno:1988bi}. The calculated curve is similar as those
given in \cite{Kouno:1988bi,Bugaev:1988dy,Singh:2009jd}.

\begin{figure}[ht]
 \centering
 \includegraphics[width=0.44\textwidth]{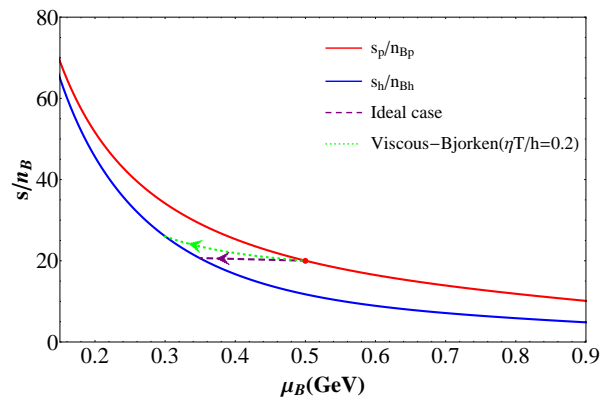}
 \caption{The ratio $s_p/n_{Bp}$ and $s_h/n_{Bh}$ along the first-order phase
 transition line. The dashed straight line and dotted curve show the trajectory of
 $s_m/n_{Bm}$ during the first-order phase transition in an ideal and viscous
 hydrodynamic expansion, respectively.
}
 \label{fig2-sn}
\end{figure}

For the later use the entropy density is given below,
\begin{equation}
\label{sph}
s_i=\frac{e_i+P_i-\mu_i n_{Bi}}{T} \,,
\end{equation}
where the subscript $i$ can be $p$ and $h$ denoting the parton and hadron phase,
respectively. We calculate $s_p/n_{Bp}$ and $s_h/n_{Bh}$ along the first-order
phase transition line. The results are shown in Fig. \ref{fig2-sn}.
We see that $s_p/n_{Bp}$ is larger than $s_h/n_{Bh}$ and both are decreasing
with increasing $\mu_B$.

\section{Viscous effects during the first-order phase transition}
\label{sec3:viscous}
In this section we show how the QCD matter in heavy-ion collisions crosses the
first-order phase transition line. For vanishing viscosity the total entropy is
conserved. With the conservation of the net baryon number the ratio of the entropy
density over the net baryon number density is conserved too. This means that
$s_h/n_{Bh}$ at the end of the phase transition should be equal to $s_p/n_{Bp}$
at the beginning of the phase transition. From Fig. \ref{fig2-sn} we realize that
$s_p/n_{Bp}$ is always larger than $s_h/n_{Bh}$ at any given $\mu_B$. Therefore,
$\mu_B$ (and $T$) cannot keep constant during the first-order phase transition.
Since in addition $s_h/n_{Bh}$ increases with decreasing $\mu_B$, 
$\mu_B$ of the two-phase state will change and move continuously to
a smaller value along the first-order phase transition line (see the dashed line
in Fig. \ref{fig2-sn}), while $T$ will accordingly move to a larger value \cite{Lee:1987mj}.
The first-order phase transition will end up at a smaller $\mu_B$ (or a larger $T$).
In the following we will study viscous effects on the dynamical evolution of
QCD matter during the first-order phase transition.
Obviously, $\mu_B$ (or $T$) along the first-order phase transition line will end up
at even smaller (larger) value (see the dotted curve in Fig. \ref{fig2-sn}),
since more entropy will be produced due to nonzero viscosities. 

We consider an expanding system of partons, which is undergoing the 
first-order confinement phase transition and hadronizing. There should be
a clear spatial separation between the parton and hadron phase.
Hadron bubbles will be formed. However, the description of the nucleation
process is still a challenging issue \cite{Csernai:1992tj,Gao:2016hks}.
We can imagine that
due to the statistical nature, some of the bubbles are disappearing, while the
others are growing and merging until the hadronization is complete. In this article
we get around the fluctuating bubble picture in the nucleation process and describe
the hadronization on an ensemble average. To this end we assume that each volume 
element, no matter how small it is, contains separated parton and hadron volumes.
Suppose $V$ is the volume of an expanding element in its local rest frame at
proper time $\tau$. We denote $V_p$ and $V_h$ as the volume of the parton
and hadron phase, respectively. The fraction of the parton phase is then 
$f_p=V_p/V=V_p/(V_p+V_h)$. The time dependence of $f_p$ describes the
hadronization on an ensemble average. The main conclusion of our study that we
will present now shows that the effect of nonzero viscosity is to accelerate the
decrease of $\mu_B$ and to slow down the first-order phase transition.

The energy density and the net baryon number density of the two-phase system
in the local rest frame of the considered expanding volume element are
\begin{eqnarray}
\label{em}
&& e_m = e_p f_p + e_h (1-f_p) \,, \\
\label{nm}
&& n_{Bm}=n_{Bp} f_p + n_{Bh} (1-f_p) \,,
\end{eqnarray}
where $e_p$, $e_h$, $n_{Bp}$, and $n_{Bh}$ are functions of $\mu_B$. (Corresponding
$T$ are determined by the first-order phase transition line shown in Fig. \ref{fig1-phase}.)
$f_p$ and $\mu_B$ are changing with time. Since $e_m$ and $n_{Bm}$ can be solved
from the hydrodynamic equations according to the energy and net baryon
number conservation, we can determine $f_p$ and $\mu_B$ at each time point.
The viscosity affects the time evolution of $e_m$ and, thus, affects $f_p$ 
and $\mu_B$ too. The following are the details for determining $f_p$ and
$\mu_B$.

Taking time derivative of Eq. (\ref{em}) gives 
\begin{equation}
\label{emtau}
\frac{\partial e_m}{\partial \tau}=(e_p-e_h) \frac{d f_p}{d \tau}
+ \left [ \frac{de_p}{d\mu_B} f_p +  \frac{de_h}{d\mu_B} (1-f_p) \right ]
\frac{\partial \mu_B}{\partial \tau}  \,.
\end{equation}
The left-hand side of the above equation can be obtained from the
hydrodynamic equation for the energy density \cite{Muronga:2003ta,Molnar:2009tx} 
\begin{equation}
\label{vhydro}
De_m=-(e_m+P_c+\Pi_m)\nabla_\mu U^\mu +\pi^{\mu\nu}_m \nabla_{<\mu} U_{\nu>}\,,
\end{equation}
where $U^\mu$ is the fluid four-velocity, $\Pi_m=\Pi_pf_p+\Pi_h(1-f_p)$ is the total
bulk pressure, and $\pi^{\mu \nu}_m=\pi^{\mu \nu}_p f_p+\pi^{\mu \nu}_h (1-f_p)$ is 
the total shear tensor. Since we use the Landau's definition of the fluid four-velocity,
there is no heat flow term in Eq. (\ref{vhydro}). Other symbols in this equation are
defined as follows:
\begin{eqnarray}
&&D\equiv U^\mu \partial_\mu \,, \\
&&\Delta^{\mu\nu} \equiv g^{\mu\nu}-U^\mu U^\nu \,,\\
&&\nabla^\mu \equiv \Delta^{\mu\nu} \partial_\nu \,,\\
&&A^{<\mu\nu>} \equiv \left [ \frac{1}{2} \left ( \Delta^\mu_\sigma \Delta^\nu_\tau
+\Delta^\nu_\sigma \Delta^\mu_\tau \right ) -\frac{1}{3} \Delta^{\mu\nu} \Delta_{\sigma\tau}
\right ] A^{\sigma\tau} \,.
\end{eqnarray}
We have then
\begin{eqnarray}
\label{exprate}
\nabla_\mu U^\mu&=&\partial_\mu U^{\mu}+\Gamma^{\mu}_{\alpha\mu}U^\alpha \,, \\
\label{gradient}
\nabla^{<\mu}U^{\nu>}&=&\frac{1}{2}(\partial^{\mu}U^{\nu}-U^{\mu}U^{\alpha}\partial_\alpha U^\nu
+\partial^{\nu}U^{\mu} \nonumber \\
&& -U^{\nu}U^{\alpha}\partial_\alpha U^\mu) 
+\frac{1}{2}(\Delta^{\mu\alpha}U^{\beta}\Gamma^{\nu}_{\alpha\beta} \nonumber \\
&&+\Delta^{\nu\alpha}U^{\beta}\Gamma^{\mu}_{\alpha\beta})
-\frac{1}{3} \nabla_\alpha U^\alpha \Delta^{\mu\nu} \,,
\end{eqnarray}
where  $\Gamma^{\mu}_{\alpha\beta}\equiv\frac{1}{2}g^{\mu\nu}(\partial_\beta g_{\alpha\nu}-\partial_\alpha g_{\nu\beta}-\partial_\nu g_{\alpha\beta})$
denotes the Christoffel symbol. By introducing the shear pressure
\begin{equation}
\label{shearpressure}
\tilde \pi_m=- \frac{\pi_m^{\mu\nu} \nabla_{<\mu} U_{\nu>}}{\nabla_\mu U^\mu} \,,
\end{equation}
Eq. (\ref{vhydro}) changes to
\begin{equation}
\label{vhydro2}
De_m=-(e_m+P_c+\Pi_m+\tilde \pi_m)\nabla_\mu U^\mu \,.
\end{equation}
In the local rest frame, where $U^\mu=(1,0,0,0)$, we have
$De_m=\partial e_m/\partial \tau$ and
\begin{equation}
\nabla_\mu U^\mu=\frac{1}{V} \frac{dV}{d\tau} \,.
\end{equation}
By equating the right-hand side of both Eqs. (\ref{emtau}) and (\ref{vhydro2}) 
we obtain
\begin{eqnarray}
\label{fp1}
\frac{d f_p}{d \tau}&=&-\frac{e_m+P_c+\Pi_m+\tilde \pi_m}{e_p-e_h} 
\frac{1}{V} \frac{dV}{d\tau} \nonumber \\
&& -\frac{1}{e_p-e_h}  \left [ \frac{de_p}{d\mu_B} f_p +  \frac{de_h}{d\mu_B} (1-f_p) \right ]
\frac{\partial \mu_B}{\partial \tau} \,.
\end{eqnarray}

Analogously to the derivation from Eqs. (\ref{em}) and (\ref{vhydro}) to Eq. (\ref{fp1}),
we can also derive $d f_p/d \tau$ from the net baryon number density 
(\ref{nm}) and its hydrodynamic evolution
\begin{equation}
\label{nmtau}
D n_{Bm}=-n_{Bm} \nabla_\mu U^\mu \,,
\end{equation}  
which indicates the net baryon number conservation. In the present study we have
neglected the diffusion current induced by the heat conduction. We have then
\begin{eqnarray}
\label{fp2}
\frac{d f_p}{d \tau}&=&-\left ( \frac{n_{Bh}}{n_{Bp}-n_{Bh}} +f_p \right ) 
\frac{1}{V} \frac{dV}{d\tau} -\frac{1}{n_{Bp}-n_{Bh}}  \nonumber \\
&& \times \left [\frac{dn_{Bp}}{d\mu_B} f_p +  
\frac{dn_{Bh}}{d\mu_B} (1-f_p) \right ]
\frac{\partial \mu_B}{\partial \tau} \,.
\end{eqnarray}

Equating Eq. (\ref{fp1}) and Eq. (\ref{fp2}) gives
\begin{equation}
\label{shearpressure1}
\Pi_m+\tilde \pi_m=\frac{n_{Bh}(e_p-e_h)}{n_{Bp}-n_{Bh}}
-(e_h+P_c) 
+ C_1 \frac{1}{\frac{1}{V} \frac{dV}{d\tau}} \frac{\partial \mu_B}{\partial \tau}  \,,
\end{equation}
where
\begin{eqnarray}
C_1&=&\left ( \frac{e_p-e_h}{n_{Bp} -n_{Bh}}
\frac{dn_{Bp}}{d\mu_B} - \frac{de_p}{d\mu_B}  \right ) f_p \nonumber \\
&&+  \left ( \frac{e_p-e_h}{n_{Bp} -n_{Bh}}
\frac{dn_{Bh}}{d\mu_B} - \frac{de_h}{d\mu_B}  \right ) (1-f_p) \,.
\end{eqnarray}
In the first-order theory of hydrodynamics, the bulk pressure and shear stress tensor are
proportional to the bulk and shear viscosity 
\cite{Muronga:2003ta,Eckart:1940te,Denicol:2014vaa},
\begin{eqnarray}
\label{Pim}
\Pi_m &=& -\xi_m \nabla_\mu U^\mu \,,\\
\label{pimmunu}
\pi_m^{\mu\nu}&=&2\eta_m \nabla^{<\mu}U^{\nu>} \,.
\end{eqnarray}
$\eta_m$ and $\xi_m$ are the shear and bulk viscosity of the two-phase system and
$\eta_m=\eta_p f_p + \eta_h (1-f_p)$ and $\xi_m=\xi_p f_p + \xi_h (1-f_p)$,
where $\eta_p$ and $\eta_h$ ($\xi_p$ and $\xi_h$) are the shear (bulk) viscosity of
the parton and hadron phase respectively.
If all the viscosities and the fluid velocity are known, we can solve $\mu_B(\tau)$
from Eq. (\ref{shearpressure1}) and then $f_p(\tau)$ from
Eq. (\ref{fp1}) or Eq. (\ref{fp2}). 

As stated at the beginning of this section, $\mu_B$ should decrease along the
first-order phase transition line. However, from Eq. (\ref{shearpressure1}) it is not
obvious that $\partial \mu_B/\partial \tau$ is negative even for vanishing viscosities.
We now look at the entropy of the two-phase system, which is 
\begin{equation}
s_m\equiv \frac{e_m+P_c -\mu_B n_{Bm}}{T}=s_p f_p + s_h (1-f_p)
\end{equation}
according to Eqs. (\ref{em}) and (\ref{nm}), and the definition (\ref{sph}).
The time evolution of $s_m$ is obtained from the time evolution of $e_m$ and
$n_{Bm}$, namely Eqs. (\ref{vhydro}) and (\ref{nmtau}). We have
\begin{equation}
\label{smtau}
\frac{\partial s_m}{\partial \tau} =-s_m \nabla_\mu U^\mu + 
\frac{\Pi_m^2}{\xi_m T} + \frac{\pi_{m,\mu\nu} \pi_m^{\mu\nu}}{2\eta_m T}
\end{equation}
for the first-order viscous hydrodynamics. Analogously to the derivation of Eq. (\ref{fp1})
we get
\begin{eqnarray}
\label{fp3}
\frac{d f_p}{d \tau}&=&-\frac{T s_m+\Pi_m+\tilde \pi_m}{T(s_p-s_h)} 
\frac{1}{V} \frac{dV}{d\tau} \nonumber \\
&& -\frac{1}{s_p-s_h}  \left [ \frac{ds_p}{d\mu_B} f_p +  \frac{ds_h}{d\mu_B} (1-f_p) \right ]
\frac{\partial \mu_B}{\partial \tau} \,.
\end{eqnarray}
Equating Eq. (\ref{fp3}) with Eq. (\ref{fp2}) gives then
\begin{eqnarray}
\label{shearpressure2}
\Pi_m+\tilde \pi_m &=& T \left ( \frac{s_p}{n_{Bp}} -\frac{s_h}{n_{Bh}}
\right ) \frac{n_{Bp} n_{Bh}}{n_{Bp} -n_{Bh}} \nonumber \\
&&+ C_2 \frac{T}{\frac{1}{V} \frac{dV}{d\tau}} \frac{\partial \mu_B}{\partial \tau}  \,,
\end{eqnarray}
where
\begin{eqnarray}
C_2&=&\left ( \frac{s_p-s_h}{n_{Bp} -n_{Bh}}
- \frac{ds_p}{dn_{Bp}}  \right ) \frac{dn_{Bp}}{d\mu_B}  f_p \nonumber \\
&&+  \left ( \frac{s_p-s_h}{n_{Bp} -n_{Bh}}
 - \frac{ds_h}{dn_{Bh}}  \right ) \frac{dn_{Bh}}{d\mu_B}(1-f_p) \,.
\end{eqnarray}
With the Gibbs-Duhem equation $dP=sdT+n d\mu$ one can prove that
Eqs. (\ref{shearpressure1}) and (\ref{shearpressure2}) are identical. 
Since both $s_p/n_{Bp}$ and $s_h/n_{Bh}$ decrease with increasing $\mu_B$
(see Fig. \ref{fig2-sn}), i.e., $d(s_p/n_{Bp})/d\mu_B < 0$ and 
$d(s_h/n_{Bh})/d\mu_B < 0$, one obtains easily $ds_p/dn_{Bp} < s_p/n_{Bp}$
and $ds_h/dn_{Bh} < s_h/n_{Bh}$. We have then
\begin{eqnarray}
\label{c2}
C_2 &>& \left ( \frac{s_p}{n_{Bp}} - \frac{s_h}{n_{Bh}}  \right )
\left [ \frac{n_{Bh}}{n_{Bp}-n_{Bh}} \frac{dn_{Bp}}{d\mu_B}  f_p \right .\nonumber \\
&& \left . +  \frac{n_{Bp}}{n_{Bp}-n_{Bh}} \frac{dn_{Bh}}{d\mu_B}(1-f_p) 
\right ] \nonumber \\
&>& 0
\end{eqnarray}
for $s_p/n_{Bp} > s_h/n_{Bh}$ (see Fig. \ref{fig2-sn}).
From Eq. (\ref{shearpressure2}) we realize that $\partial \mu_B/\partial \tau$ is
always negative and its absolute value becomes larger for increasing
viscosity. [Remember that $\Pi_m$ and $\tilde \pi_m$ are negative for the
first-order viscous hydrodynamics when the system is expanding and they are
proportional to the bulk and shear viscosity according to
Eqs. (\ref{shearpressure}), (\ref{Pim}), and (\ref{pimmunu}).] The viscous effect
leads to a stronger decrease of $\mu_B$
during the first-order phase transition, compared to the ideal hydrodynamic
expansion. Moreover, according to Eq. (\ref{fp2}) the decrease of $f_p$ slows
down in the viscous case. The larger the viscosity, the longer will the phase
transition take.

For demonstrating the viscous effects we now calculate explicitly
the time evolution of $\mu_B$ and $f_p$ during the first-order phase transition
with given fluid velocity and viscosities. We compare the results with 
nonzero viscosities to those with zero viscosities.

To this end we use the analytical solutions of $U^\mu$ from one-dimensional
Bjorken expansion \cite{Bjorken:1982qr} and three-dimensional Gubser expansion
\cite{Gubser:2010ze,Gubser:2010ui}.
Be $U^\mu=\gamma(1,{\bf v})$ with $\gamma=1/\sqrt{1-v^2}$ in the space time
coordinate $(t,{\bf r})$. With the time $\tilde \tau=\sqrt{t^2-z^2}$, the
space time rapidity $\eta=(1/2)\ln(t+z)/(t-z)$, the transverse radius
$\rho=\sqrt{x^2+y^2}$ and the azimuthal angle $\phi$, the fluid velocity
can be transformed into the coordinate $(\tilde \tau, \eta, \rho, \phi)$ as follows
\cite{Liao:2009zg}:
\begin{eqnarray}
U^{\tilde \tau}&=&\gamma(\cosh\eta-v_z\sinh\eta) \,, \nonumber \\
U^\eta&=&\frac{\gamma}{\tilde \tau}(v_z\cosh\eta-\sinh\eta) \,, \nonumber \\
U^\rho&=&\gamma(v_x\cos\phi+v_y\sin\phi) \,, \nonumber \\
U^\phi&=&\gamma(v_y\cos\phi-v_x\sin\phi) \,.
\end{eqnarray}
$U^\mu$ in Bjorken expansion is given in the coordinate $(t,{\bf r})$
 \cite{Bjorken:1982qr},
\begin{equation}
v_x=v_y=0 \,, \ v_z=\frac{z}{t} \,,
\end{equation}
while $U^\mu$ in Gubser expansion is given in the coordinate
$(\tilde \tau, \eta, \rho, \phi)$ \cite{Gubser:2010ze,Gubser:2010ui},
\begin{equation}
U^{\tilde \tau}=\cosh k \,, \ U^\rho=\sinh k \,, \ U^\eta=U^\phi=0 \,,
\end{equation}
where
\begin{equation}
\tanh k=\frac{2\tilde \tau \rho}{a^2+\tilde \tau^2+\rho^2} \,.
\end{equation}
Different from the Bjorken expansion, the Gubser expansion includes transverse
expansion. The parameter $a$ is set to be $4.5 \mbox{ fm}$.
A similar value has been used to describe the hydrodynamic evolution of QGP in 
Au+Au collisions at RHIC with $\sqrt{s_{NN}}=200 \mbox{ GeV}$ \cite{Gubser:2010ze,Kolb:2002ve}.
In addition we choose $\rho=0$ in our calculations. The phase transition in 
volume elements with larger transverse radius $\rho$ will occur earlier. 
For Bjorken expansion and for Gubser expansion at $\rho=0$, $\tilde \tau$ is equal
to the proper time in the local rest frame, $\tau$.

Although there are calculations and model-to-data analyses on the shear
and bulk viscosity of the parton and/or hadron phase in heavy-ion collisions
\cite{Demir:2008tr,Sasaki:2008fg,Khvorostukhin:2010aj,Wesp:2011yy,NoronhaHostler:2012ug,Ozvenchuk:2012kh,Ryu:2017qzn,Auvinen:2017fjw}, 
the shear and bulk viscosity of both phases during the first-order phase
transition are not yet fixed so far. 
We assume {\it for simplicity}  that the shear and bulk viscosity in the
parton phase are equal and the shear and bulk viscosity in the hadron phase are
twice as much as those in the parton phase. Moreover, we set 
$\eta_pT/h_p=\xi_pT/h_p$ to be constant during the phase transition. 
$h_p=e_p+P_c$ is the enthalpy density of partons.
At $\mu_B=0$, $h_p/T$ is equal to the entropy density. $\eta_p T/h_p$ and 
$\xi_p T/h_p$ are more relevant to characterize the viscous effect in slow
expansion \cite{Liao:2009gb} as will happen in heavy-ion collisions with lower
colliding energies.

\begin{figure}[ht]
 \centering
 \includegraphics[width=0.45\textwidth]{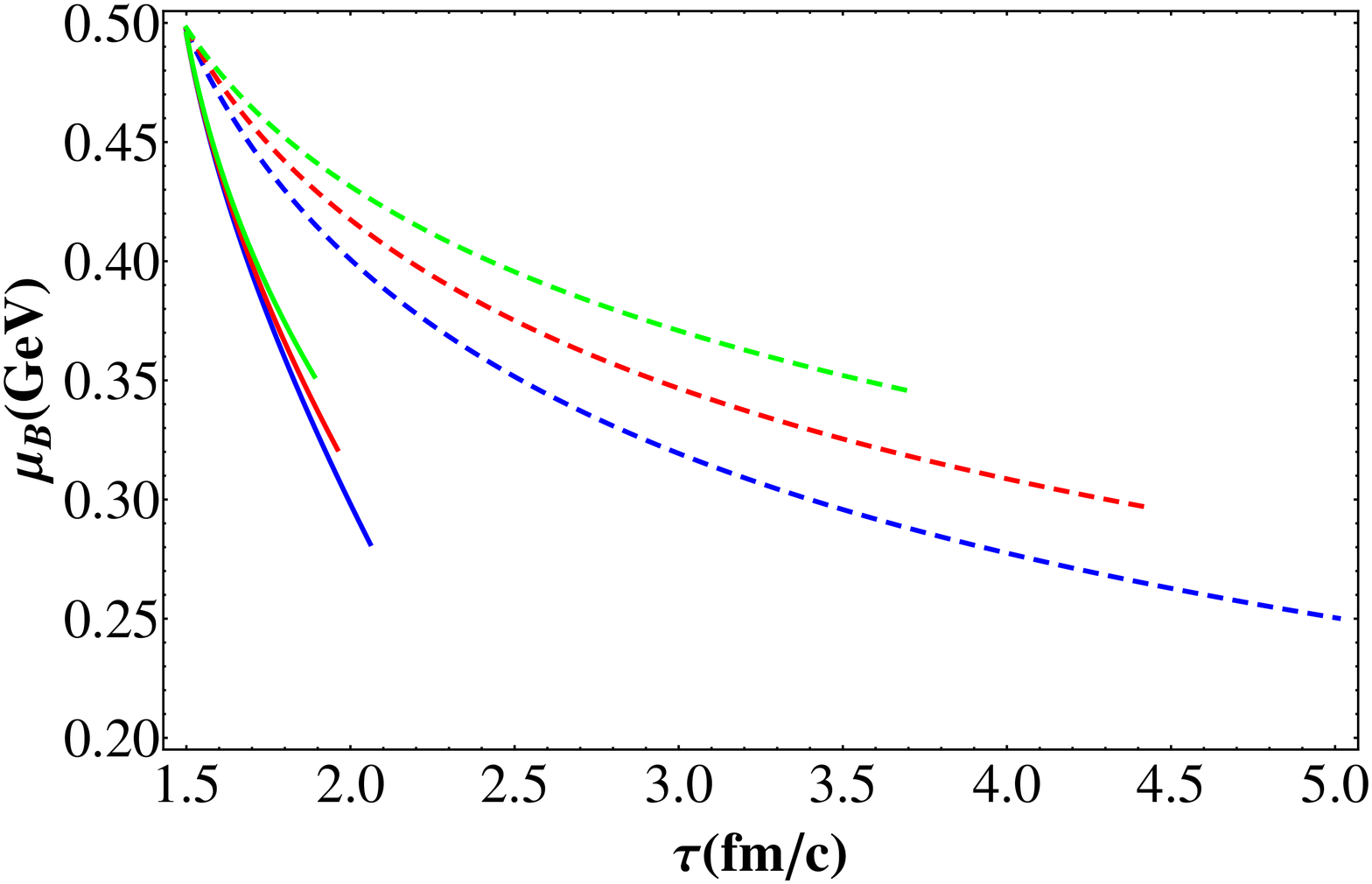}
 \includegraphics[width=0.45\textwidth]{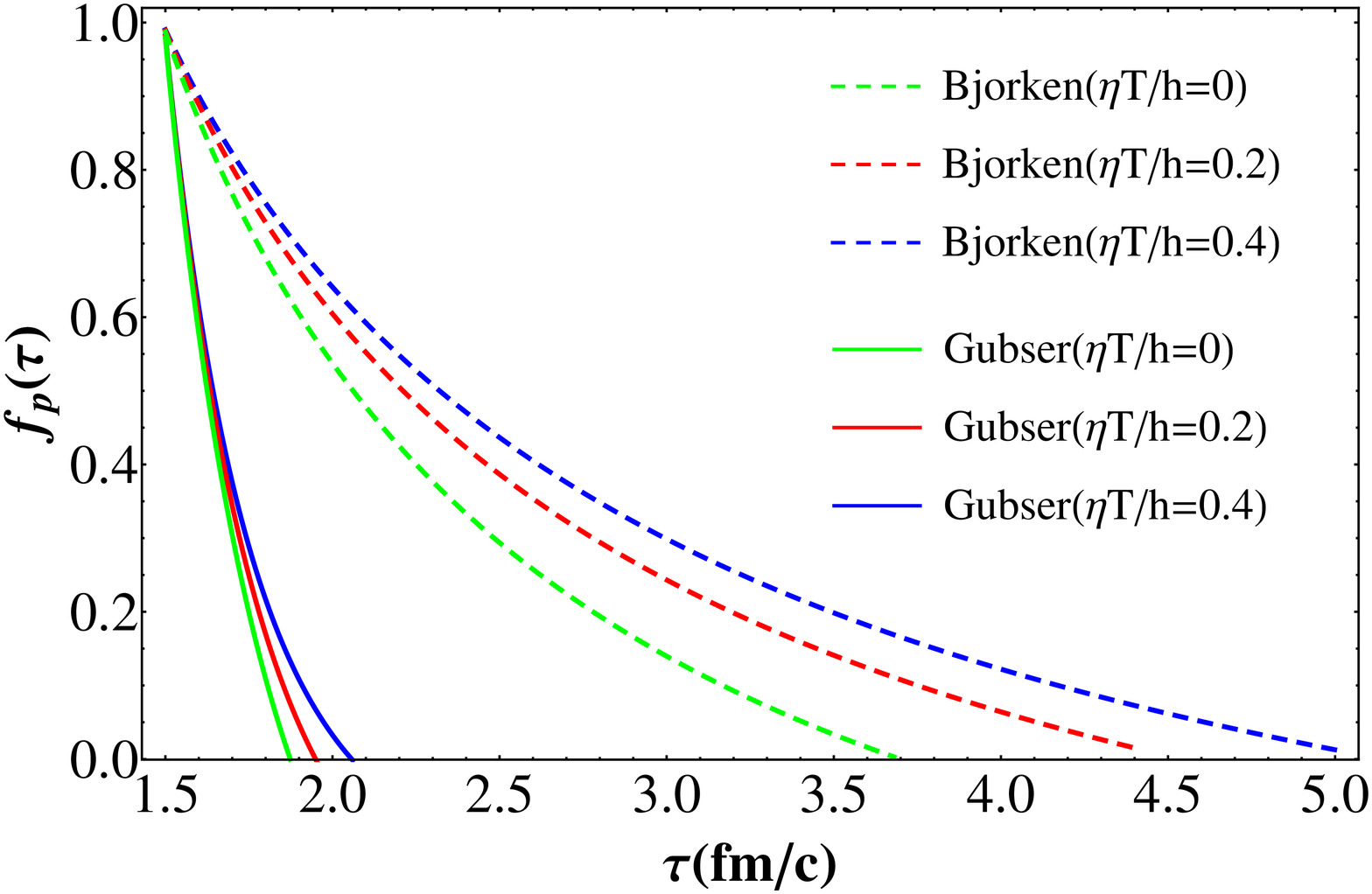}
 \includegraphics[width=0.45\textwidth]{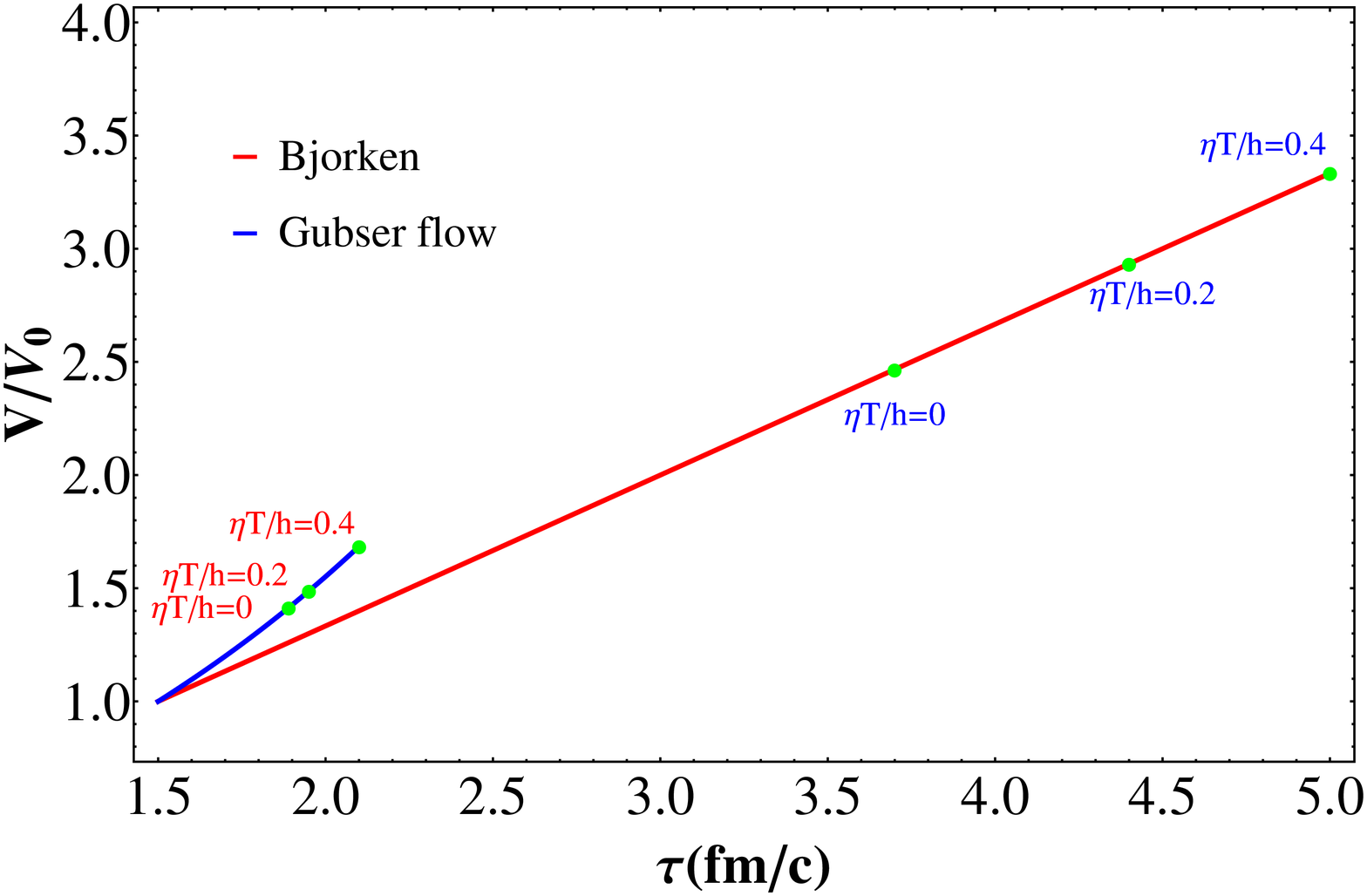}
 \caption{Time evolution of $\mu_B$, $f_p$, and volume increase for ideal
 and viscous hydrodynamic expansion. We set $\tau_c=1.5 \mbox{ fm/c}$
 and $\mu_B(\tau_c)=0.5 \mbox{ GeV}$.
}
 \label{fig3-muBfp}
\end{figure}

In Fig. \ref{fig3-muBfp} the time evolution of $\mu_B$ and $f_p$ are presented
with three different values of viscosities, $\eta_pT/h_p=\xi_pT/h_p=0, 0.2, 0.4$,
and with two different expansion dynamics, Bjorken and Gubser expansion.
As an example, the starting time of the phase transition $\tau_c$ was set to 
be $1.5 \mbox{ fm/c}$ and $\mu_B$ at $\tau_c$ has been chosen to be $0.5 \mbox{ GeV}$.
With these settings we obtain $n_{Bp}=0.1948/ \mbox{fm}^3$ and $s_p/n_{Bp}=20$
at $\tau_c$.

First we see that the results agree with the qualitative analyses done before.
Compared with those with zero viscosities, the results with increasing viscosities 
show that the decrease of $\mu_B$ becomes stronger, the phase transition takes
longer, and thus $\mu_B$ ends up with smaller final values, when the phase transition
in the considered volume is complete.
Second, the final value of $\mu_B$ in ideal hydrodynamic expansion does not depend on
the expansion geometry, because both the entropy and net baryon number are conserved.
In viscous expansion the entropy production depends on the expansion geometry. Therefore,
the final value of $\mu_B$ in the Bjorken expansion is different from that in the Gubser
expansion. Third, compared to the Bjorken expansion, the transverse expansion in the
Gubser expansion leads to a stronger decrease of $\mu_B$ and faster phase transition.

In addition, we show in the last panel of Fig. \ref{fig3-muBfp} the volume increase
during the phase transition, $V(\tau)/V(\tau_c)$. The points mark the volume increase 
at different end times of the phase transition. The rate of the volume increase is
stronger in three-dimensional Gubser expansion than in one-dimensional Bjorken expansion. 
However, since the phase transition proceeds faster in Gubser expansion than in
Bjorken expansion, the volume increase at the end of the phase transition is 
stronger in Bjorken expansion than in Gubser expansion.

The trajectory of $s_m/n_{Bm}$ during the first-order phase transition is plotted 
in Fig. \ref{fig2-sn} for an ideal expansion (dashed straight line) and a viscous 
Bjorken expansion with $\eta_p T/h_p=0.2$ (dotted curve).

Our results of the viscous effects on the dynamical evolution of QCD matter during
the first-order phase transition
could be refined, if more reliable EoS and transport coefficients of the parton and hadron
phase were available and more realistic hydrodynamic expansion of QCD matter 
had been calculated at large baryon chemical potential. The present study is the basis
for a further development of the dynamic transport simulation of the QCD phase transition
in heavy-ion collisions \cite{Feng:2016ddr}.

\section{Further Discussions}
\label{sec4:lim}
In this section we discuss (or speculate) how the first-order phase transition
will proceed with much larger viscosities, with which calculations using the
first-order hydrodynamics may be invalid.

Recalling Eq. (\ref{emtau}) at the starting time of the phase transition $\tau_c$,
where $f_p=1$, we have
\begin{equation}
\label{emtauc}
\frac{\partial e_m}{\partial \tau}=(e_p-e_h) \frac{d f_p}{d \tau}
+ \frac{de_p}{d\mu_B} \frac{\partial \mu_B}{\partial \tau}  \,.
\end{equation}
For an expanding system $\partial e_m/\partial \tau$ is always negative, while
its absolute value depends on the form of the expansion ($U^\mu$) and viscosity.
The larger the viscosity, the smaller
is $|\partial e_m/\partial \tau|$. If the second term on the right hand side of 
Eq. (\ref{emtauc}), $(de_p/d\mu_B)(\partial \mu_B/\partial \tau)$, 
is positive, $df_p/d\tau$ should be negative, which indicates that the
first-order phase transition will always proceed with any large viscosity, unless
the viscous hydrodynamics breaks down.
On the other hand, if $(de_p/d\mu_B)(\partial \mu_B/\partial \tau)$
is negative, $df_p/d\tau$ could be (mathematically) positive for sufficient large
viscosity and slow expansion. A positive $df_p/d\tau$ at $\tau_c$ is not
physical, which indicates that large viscosity may forbid the occurrence of
first-order phase transition. 

Both the signs of $de_p/d\mu_B$
and $\partial \mu_B/\partial \tau$ are determined by the EoS of the parton and
hadron phase. For the EoS used in this work, both $de_p/d\mu_B$ and
$\partial \mu_B/\partial \tau$ are negative. Thus, $df_p/d\tau$ at $\tau_c$
is always negative.

We have to note that for large viscosity the friction heat will be so large, that 
$\partial e_m/\partial \tau$ becomes positive when using the first-order hydrodynamics
[see Eqs. (\ref{Pim}), (\ref{pimmunu}), and (\ref{vhydro})]. A positive 
$\partial e_m/\partial \tau$ can lead to a positive $df_p/d\tau$ at $\tau_c$
according to Eq. (\ref{emtauc}). However, positive $\partial e_m/\partial \tau$ is
impossible for an expanding system. In this case the first-order
hydrodynamics is invalid and higher order terms have to be included in the hydrodynamic 
description of the phase transition.

The phase transition for $\tau > \tau_c$ seems more complicated. It cannot
be proven from 
Eq. (\ref{emtau}) that $df_p/d\tau$ is always negative with the used EoS, 
because $de_h/d\mu_B$ is positive. Thus, with sufficient large viscosity
and slow expansion (slower than Bjorken and Gubser expansion), $df_p/d\tau$
may become positive and a transition of the
net baryon number from the hadron phase to the parton phase may happen.
However, this will not lead to the disappearance of the hadron phase, since
$df_p/d\tau$ will be negative again at least at $f_p=1$, as proven before. 
Thus, we expect that for large viscosity and slow expansion, the time evolution
of $f_p$ trends to decrease from $1$ to $0$, but maybe has some humps in
between.

We note that at $\mu_B$ being smaller than the value at the critical point,
the QCD phase transition is a crossover. Our formalism developed in the previous
section does not work in the crossover region. Although the exact position of
the critical point connecting the crossover and first-order phase transition line
is not known yet \cite{Asakawa:1989bq,Stephanov:2004wx,Kumar:2012fb,Luo:2015doi}, 
various theoretical calculations suggest that its most probable location is in
the $\mu_B$ interval between 200 and 400 MeV \cite{Fodor:2004nz,Gavai:2004sd}.
For large viscosity, the moving $(\mu_B, T)$ point along the first-order phase
transition line may pass the critical point. Then critical phenomena are
expected to occur.

The whole derivations in the previous section from Eq. (\ref{em}) to Eq. (\ref{c2})
are also valid for the first-order phase transition in a contracting medium,
where $dV/d\tau$ is negative.
In this case, $\mu_B$ increases with time along the first-order phase transition
line. We then describe a transition from the hadron phase to the parton phase.

In heavy-ion collisions the produced QCD matter expands on the whole. But
locally on a small spatial scale, expansion as well as contraction exist due
to density fluctuations. Our study provides a potential hydrodynamic framework 
to describe the nucleation of partons.

\section*{Acknowledgement}
The authors are grateful to U. Heinz for reading the manuscript and very valuable comments.
Z.X. thanks P. Huovinen, Y. Liu, N. Xu, and P. Zhuang for helpful discussions.
This work was financially supported by the National Natural Science
Foundation of China under Grants No. 11575092, No. 11335005, and No. 11275103,
and the Major State Basic Research Development Program in China under Grants
No. 2015CB856903. C.G. acknowledges support by the Deutsche
Forschungsgemeinschaft (DFG) through the grant CRC-TR 211
``Strong-interaction matter under extreme conditions''.

\end{document}